# Measuring the Novelty of Biomedical Papers Using the Latent Distances between Knowledge Units

Yi Zhao[1], Heng Zhang[2], Yuzhuo Wang[1], Wenqing Wu[3], Tong Bao[3], Chengzhi Zhang[3*]

1. School of Management, Anhui University, Hefei, China
2. School of Information Management, Central China Normal University, Wuhan, China
3. Department of Information Management, Nanjing University of Science and Technology, Nanjing, China

**Abstract:** Measuring the novelty of scientific papers is a central concern in research evaluation and scientometrics. From a recombination perspective, prior studies have largely focused on the co-occurrence of knowledge units to assess the novelty of scientific papers. However, these studies often overlook other relationships between knowledge units. This narrow view may result in inaccurate or incomplete evaluations of novelty for scientific papers. To fill this gap, this study introduces a comprehensive novelty measurement that incorporates three types of relationships between knowledge units: network, semantic, and hierarchical. These relationships are used to quantify the latent distances among knowledge units. Using a dataset of 142,036 articles published in *PLoS ONE* and a validation dataset from the H1 Connect platform, our results demonstrate that (1) each relationship type captures distinct latent distances between MeSH terms; (2) compared to the widely used indicators proposed by Uzzi et al. (2013), our measures show stronger alignment with peer judgements; and (3) combining all three distance metrics yields more effective identification of novel papers than using any single perspective alone.

# 1. Introduction

Novelty stands as a crucial criterion in the innovative evaluation of academic papers, and its measurement has attracted significant attention from the scholarly community [1, 2]. Although

* Corresponding author

*Email addresses*: yizhao93@ahu.edu.cn (Yi Zhao), zh_heng@ccnu.edu.cn (Heng Zhang), wangyuzhuo@ahu.edu.cn (Yuzhuo Wang), winchywwq@njust.edu.cn (Wenqing Wu), tbao@njust.edu.cn (Tong Bao), zhangcz@njust.edu.cn (Chengzhi Zhang)

peer review remains the gold standard for assessing scientific novelty, the process itself has drawn substantial criticism. First, peer reviewers' conservative tendencies frequently impede the fair assessment of highly novel work [3]. Second, despite the rapid growth in scholarly articles, the pool of available reviewers has not expanded proportionally, which may compromise the quality of peer assessment [4]. Third, peer review is inherently subjective as a method for assessing novelty, and its outcomes often lack consistency due to reviewers' cognitive biases and varying expertise [5]. To overcome these limitations, several efforts have been devoted to developing automated measures of scientific novelty for academic publications.

Scholars argue that novelty does not emerge from ex nihilo but rather from the recombination of existing knowledge units [6, 7]. In studies measuring novelty, knowledge units are typically operationalized through various sources, including referenced journals [6, 8], keywords [9], knowledge entities [10, 11], and sentences [12]. Although prevailing studies have used the unusualness of journal pairs in reference lists to quantify scientific novelty, some scholars argue that referenced journals are not an appropriate source for capturing fine-grained scientific knowledge [4, 12]. A distinct line of criticism further contends that such reference-based measures may reflect interdisciplinarity rather than scientific novelty [13]. These limitations collectively underscore the need for developing content-based approaches to novelty measurements.

While several studies have proposed novelty indicators based on paper titles, titles may not fully capture the knowledge contained in scientific publications [1, 12]. Consequently, scholars have increasingly used keywords or knowledge entities as proxies for knowledge units [14]. Existing novelty measures based on these elements primarily rely on co-occurrence relationships. However, knowledge units may also be related through semantic relationships and hierarchical structures, suggesting that co-occurrence alone may not fully capture the relationships among knowledge units and may therefore provide an incomplete assessment of scientific novelty [15]. For instance, in the deep learning era, Convolutional Neural Network (CNNs) became a key approach for image classification tasks due to their exceptional feature extraction capabilities [16].

Prior studies have examined knowledge combinations from three perspectives: co-occurrence, semantic relationships, and hierarchical structures [9, 17, 18]. The co-occurrence perspective defines relationships based on the appearance of distinct knowledge units within the same publication [9]. The semantic perspective captures the conceptual distance between knowledge units through semantic analysis [17]. The hierarchical perspective draws on the structure of knowledge classification systems, reflecting domain-specific organization and expert knowledge [18]. These perspectives capture different aspects of relationships among knowledge units and may provide complementary information for novelty measurement.

Nevertheless, existing novelty indicators typically focus on only one perspective and rarely integrate all three. As a result, current approaches may not fully represent the complexity of knowledge combinations in scientific research.

To this end, we propose a novel and comprehensive method for measuring the novelty of academic papers. Using MeSH (Medical Subject Headings) terms as proxies for knowledge units, we define novelty as the atypical combination of prior knowledge (i.e., MeSH terms). Our approach integrates three distance metrics, including network, semantic, and hierarchical, to estimate the distance between MeSH term pairs. We applied our novelty measures to 142,036 papers from *PLoS One* and demonstrated that the three perspectives capture distinct aspects of distance in knowledge combinations. To validate the method, we compared our results with faculty-assessed novelty from the H1 Connect platform and benchmarked our measure against Uzzi et al.'s [6] pioneering approach, confirming its superiority.

# 2. Related work

## 2.1 Knowledge recombination as a source of scientific novelty

Schumpeter, a pioneer in innovation studies, posits that "innovation combines factors in a new way, or that it consists in carrying out new combinations" [19]. Translating this combinatorial logic into the cognitive realm, the creation of scientific novelty is fundamentally driven by cognitive search and recombination processes [20]. In this context, the novelty potential of scientific research derives from the cognitive distance bridged between diverse knowledge domains or conceptual bases[21]. Heterogeneity in scholars' developmental trajectories and training environments leads to differences in cognitive frameworks, resulting in variation in how they interpret, understand, and evaluate external knowledge and phenomena[22]. Consequently, cognitive distance refers to the divergence in cognitive frames or knowledge space that determines how readily actors can comprehend and integrate each other's knowledge[23].

In the complex landscape of science, knowledge is not organized along a single form, but is simultaneously structured relationally, conceptually, and taxonomically [18, 24, 25]. Such organizational forms jointly shape the structure of the knowledge space and influence scholars' perceptions of the distance between knowledge elements, thereby affecting the novelty of scientific discoveries. This can be attributed to the recombinant nature of scientific discovery and the associated knowledge search processes that underlie it [26]. First, co-occurrence knowledge networks provide a relational representation of knowledge structures by transforming patterns of joint appearance among entities into a network of associations that reflect the collective knowledge of scientific communities. Distance in this network operationalizes relational separation or structural search scope, and the recombination of knowledge units at large network distances resembles a breadth-first search process for identifying novel opportunities for knowledge recombination [27]. Second, semantic distance measures conceptual coherence and determines whether structurally distant units can be cognitively integrated into a meaningful configuration. Even when structural links exist, excessive semantic mismatch may hinder successful conceptual blending[25]. Third, a taxonomy is a hierarchical classification schema that organizes entities based on either their intrinsic properties or their relationships within a domain, and it typically reflects the structured knowledge and cognitive judgments of domain experts [18]. Distance within a knowledge taxonomy captures path-based separation in controlled taxonomies, encompassing both vertical movements across levels of abstraction and horizontal movements across branches. Such distance is associated with cognitive reconfiguration in expert knowledge representations. Collectively, these three distances act as interrelated cognitive filters shaping the knowledge search and knowledge combination process, suggesting that the novelty of scientific publications should be understood through multiple dimensions of cognitive distance.

From a practical perspective, recent studies in knowledge representation and downstream prediction tasks have demonstrated the effectiveness of integrating multiple heterogeneous information sources to better capture complex relationships among knowledge elements. For instance, Ding and Jin show that incorporating corpus-based co-occurrence information, ontology-derived hierarchical structures, and semantic predications can significantly improve the quality of MeSH term representation and enhance performance in downstream tasks such as link prediction [28]. Their results indicate that different data sources provide complementary views of the knowledge space, including relational, semantic, and hierarchical information. Similarly, Liu et al. [27] propose a multi-view heterogeneous hypergraph framework for knowledge combination prediction, in which co-occurrence, co-citation, and hierarchical structure are jointly modeled to capture different aspects of knowledge interactions. Their findings further suggest that integrating multiple structural views of knowledge can improve the accuracy of predicting future knowledge combinations, as each view reflects distinct but complementary mechanisms of knowledge organization. These studies highlight that integrating multiple views of knowledge yields a more comprehensive understanding of how knowledge elements are related and recombined. This provides a strong rationale for quantifying scientific novelty through multiple dimensions of cognitive distance, as different distance measures capture complementary aspects of knowledge organization.

To rigorously quantify these distinct cognitive dimensions, a structured and formal representation of scientific knowledge is required. Scientific knowledge is objective, falsifiable, and evolutionary in nature, often organized in a hierarchical structure [29–31]. The MeSH thesaurus conforms to this characterization and is widely used as a proxy for scientific knowledge, serving as a common basis for assessing the novelty of scientific articles [32, 33]. Updated annually, MeSH thesaurus incorporates new descriptors, removes obsolete concepts, revises scope notes, and restructures hierarchical branches to reflect the evolving landscape of biomedical knowledge [34]. Because MeSH thesaurus is curated by domain experts, it ensures terminological consistency and reduces semantic ambiguity. The combination of MeSH terms assigned to scientific articles thus represents how authors strategically integrate existing bodies of knowledge and provides a meaningful indication of an article's novelty. In practice, this operational approach has been widely adopted to quantify scientific novelty [33, 35]. Overall, both theoretically and practically, the MeSH thesaurus serves as a robust and reliable foundation for capturing the novel contributions embedded in scientific articles.

## 2.2 Quantification of scientific novelty

With the advancement of technologies such as natural language processing and complex networks, some scholars have begun developing automated methods to assess the novelty of academic papers. Current approaches to measuring the scientific novelty of academic papers fall into two main categories: reference-based and content-based novelty measures.

From the combinatorial perspective of scientific progress, scientific novelty can be viewed as the recombination of prior knowledge in unprecedented ways [36, 37]. Building on this view, Uzzi et al. [6] treated referenced journals as proxies for knowledge units, quantifying novelty through a Z-score measure based on the atypicality of journal combinations in a paper's reference list. Lee and colleagues [8] subsequently refined this method by computing the expected co-occurrence probability of journal pairs within a given year, using the ratio of observed to expected co-occurrences as a novelty indicator. This method reduced computational cost compared to that of Uzzi et al. [6] method. Wang et al. [37] further advanced this method by employing historical co-occurrence data to construct a journal co-occurrence matrix and derive vector representations for journals. The authors assessed novelty through cosine similarity between journal vectors, where lower similarity indicated greater novelty.

Recently, scholars have explored using textual information, such as keywords and knowledge entities, to represent scientific knowledge. For instance, Mishra and Torvik [38] developed a novelty measure based on MeSH term combinations, incorporating term age and frequency. Yan et al. [9] quantified a paper's novelty by computing the ratio of new keyword pairs. New keyword pairs are defined as unique combinations previously unobserved in the given research field. Similarly, Liu et al. [17] adopted bio-entities as proxies for knowledge units and proposed an entity-based novelty measure, where a paper's novelty was determined by the ratio of novel bio-entity pairs identified via semantic distance thresholds. Luo et al. [10] utilized combinations of question and method entities to quantify temporal and semantic novelty of academic papers. Temporal novelty was measured by the age and frequency of entity pairs, while semantic novelty was computed by the average of one minus the highest cosine similarity between question entities, method entities, and their pairs. Ruan et al. [35] argued that MeSH terms serve as a more suitable proxy than references for assessing novelty, proposing a measure based on the ratio of observed to expected frequency of MeSH term pairs to determine scientific papers' novelty. Chen et al. [39] combined heterogeneous hypergraph learning with large language models and proposed a model named H2GLM to predict knowledge recombination, which opens a new avenue for the quantification of scientific novelty.

Although several novelty measures have been proposed, these methods still exhibit certain limitations. First, reference-based novelty measures rely on cited journals, which may not capture fine-grained knowledge, and thus fail to assess true novelty accurately [1, 13]. Second, author-provided keywords and extracted entities often consist of non-standardized terms, introducing noise into novelty measurements [35]. Third, most existing studies focus solely on the co-occurrence frequency of knowledge combinations while neglecting other vital relationships between them. Therefore, in this study, we used the MeSH terms as knowledge units. From a recombination lens, we introduced a new approach to measure the novelty of academic papers by incorporating the network distance, semantic distance, and hierarchical distance of MeSH term pairs.

# 3. Data and methodology

In this section, we describe the dataset employed in our empirical analysis. Next, we present our proposed novelty measurement for academic papers. Finally, we introduce the validation method used to assess the effectiveness of this measurement.

## 3.1 Data collection

### 3.1.1 Dataset for empirical analysis

This study collected 142,036 research articles published in *PLoS ONE* between 2007 and 2015, all stored in XML format. We parsed these files to extract metadata, including titles, DOIs, authors, volume/issue numbers, and other relevant information. Since the *PLoS ONE* dataset did not include MeSH terms, we integrated it with the PubMed Knowledge Graph (PKG) to obtain MeSH annotations for each article [40]. The linkage process involved two key steps: Firstly, we converted DOI to the PubMed ID (PMID) for each article using the ID Converter tool[1] from the U.S. National Library of Medicine (NLM). Secondly, we matched *PLoS ONE* articles with PKG entries based on their PMIDs. After processing, 142,034 articles were successfully mapped to the PKG and retained for analysis.

Following the practice of our previous study (Zhao et al., 2024), this study identified each article's discipline using the NSF subject classification system. Of the total dataset, 128,750 articles (90.65%) contained identifiable disciplinary information. As shown in Table 1, although *PLoS ONE* is a multidisciplinary journal, its publications are predominantly concentrated in three fields, including clinical medicine, biomedical research, and biology, which collectively account for 91.54% of all papers.

Table 1 The disciplinary distribution of *PLoS ONE* papers

| Discipline | # of publications | Ratio (%) |
|---|---|---|
| Clinical Medicine | 61,523 | 47.78 |
| Biomedical Research | 43,639 | 33.89 |
| Biology | 12,696 | 9.86 |
| Psychology | 3,186 | 2.47 |
| Earth and Space | 1,705 | 1.32 |
| Health | 1,209 | 0.94 |
| Physics | 1,144 | 0.89 |
| Engineering and Technology | 1,116 | 0.87 |
| Social Sciences | 1,005 | 0.78 |
| Chemistry | 641 | 0.50 |
| Professional Fields | 464 | 0.36 |
| Mathematics | 400 | 0.31 |
| Humanities | 21 | 0.02 |
| Arts | 1 | 0.00 |
| Total | 128,750 | 100 |

### 3.1.2 Dataset for measurement validation

Following the practice of previous studies [12, 43], this study evaluated the effectiveness of our proposed novelty measurement by comparing its agreement with expert assessments from the H1 Connect platform[2]. In addition, prior research has demonstrated that pioneer novelty measurement proposed by Uzzi et al. [6] outperforms those introduced by Wang et al. [37] and Lee et al. [8] in assessing scientific novelty [43]. Building on this finding, we further conducted a comparative analysis between our proposed measurement and the Uzzi et al.'s [6] approach.

H1 Connect (formerly known as F1000Prime) is a biomedical article recommendation platform where experts conduct post-publication evaluations and provide recommendations. Additionally, each recommended article was assigned one or more specific tags to imply the justification for its recommendation by the experts. The tags have nine types, as shown in Table 2. Following Bornmann et al. [43], we considered papers labeled with the tags "Technical advance", "New Finding", "Hypothesis", and "Novel Drug target" as novel papers.

[1] https://www.ncbi.nlm.nih.gov/pmc/tools/idconv/

[2] https://archive.connect.h1.co/

Table 2 Tags assigned to recommended papers in H1 Connect

| Tag | Definition |
|---|---|
| **Hypothesis** | article presents an interesting hypothesis |
| **New finding** | article presents original data, models or hypotheses |
| **Novel drug target** | article suggests new targets for drug discovery |
| **Technical advance** | article introduces a new practical/theoretical technique, or novel use of an existing technique |
| Confirmation | article validates previously published data or hypotheses |
| Good for teaching | key article in a field and/or is well written |
| Negative/null results | article has null or negative findings |
| Refutation | article disproves previously published data or hypotheses |
| Controversial | article challenges established dogma |

This study utilized web crawling to retrieve *PLoS ONE* articles endorsed by faculty members on the H1 Connect platform. We identified 2,036 faculty-recommended articles published between 2007 and 2015, including 1 perspective, 1 review, and 2,034 research articles. Since our validation approach required disciplinary classifications, issue/volume metadata, and recommendation tags, articles lacking these criteria were excluded, yielding a final dataset of 1,837 articles.

The novelty measure proposed by Uzzi et al. [6] is computationally intensive to calculate. We therefore obtained this measure (hereafter termed "novelty score U") from the SciSciNet dataset [44], which was constructed using the Microsoft Academic Graph (MAG). By linking our *PLoS ONE* dataset with SciSciNet using DOIs, we acquired precomputed novelty scores for 128,120 *PLoS ONE* articles.

## 3.2 Novelty measurement of scientific papers using latent distances between pairs of MeSH terms

MeSH is a controlled vocabulary system developed by the U.S. National Library of Medicine

for indexing life science literature [45]. The MeSH system comprises four types[3]: (1) Main Headings, which describe the primary subject of the publication; (2) Subheadings that qualify and refine Main Headings; (3) Supplementary Concept Records that complement the main vocabulary; and (4) Publication Characteristics that describe publication types or study attributes. Main Headings have a hierarchical structure and are organized into 16 categories, including Category A (Anatomy), Category B (Organisms), and others. Following previous practice [15, 28, 35, 46], this study focuses solely on Main Headings (hereafter, MeSH terms) for subsequent analysis.

From a recombination perspective, novel knowledge combinations can be viewed as distant or unusual pairings of existing knowledge units. Greater distances between knowledge units in the knowledge space indicate more novel combinations. Papers with more novel combinations exhibit higher novelty [4]. In this study, we used MeSH terms as proxies for knowledge units and quantified the novelty of MeSH combinations by comprehensively considering their network, semantic, and hierarchical distances. The detailed process for calculating the novelty of academic papers is described in this section. It is worth noting that MeSH terms were originally developed for indexing biomedical literature. Consequently, the proposed novelty metric is most suitable for applications within the biomedical domain, and its applicability to other disciplines may be limited. Nonetheless, the underlying approach is adaptable to fields that use similarly structured hierarchical classification systems, although its performance in those settings would need to be validated.

### 3.2.1 Comprehensive distances between MeSH terms

For the MeSH combination $(MeSH_{n-1}, MeSH_n)$, this study calculates three distance metrics including network distance $d_{network}$, semantic distance $d_{semantic}$, and hierarchical distance $d_{hierarchical}$. These distance metrics are then integrated into a comprehensive distance measure $d_{comprehensive}$. The calculation formula is shown in Equation (1).

$$d_{comprehensive}(MeSH_{n-1}, MeSH_n) = \alpha * d_{network}(MeSH_{n-1}, MeSH_n) + \beta * d_{semantic}(MeSH_{n-1}, MeSH_n) + (1 - \alpha - \beta) * d_{hierarchical}(MeSH_{n-1}, MeSH_n) \quad (1)$$

Where $\alpha$, $\beta$, and $1 - \alpha - \beta$ represent the weights assigned to the three distance metrics, respectively. The detailed weight calculation method is described in section 3.2.2.

**1) Network distance**

To measure the network distance between a pair of MeSH terms, denoted as $d_{network}(MeSH_{n-1}, MeSH_n)$, we first constructed a MeSH term co-occurrence network using articles from the PKG dataset [40]. In this network, each node represents a MeSH term, and two terms are connected by an edge if they co-occur in the same article. The edge weight represents their co-occurrence frequency. We then used the LINE algorithm to generate a vector representation for each MeSH term [47]. LINE preserves both local and global network structures and offers faster training speed than alternatives like DeepWalk and Node2Vec [47]. The distance was calculated as the cosine distance between embedding vectors of MeSH terms, as defined by Equation (2).

[3] https://www.nlm.nih.gov/oet/ed/pubmed/mesh/mod01/03-100.html

$$d_{network}(MeSH_{n-1}, MeSH_n) = 1 - similarity(MeSH_{n-1}, MeSH_n) = 1 - \frac{Net(MeSH_{n-1})\cdot Net(MeSH_n)}{||Net(MeSH_{n-1})||\ ||Net(MeSH_n)||} \quad (2)$$

Where the cosine similarity between MeSH terms is defined as $similarity(MeSH_{n-1}, MeSH_n)$, and $Net(MeSH_{n-1})$ and $Net(MeSH_n)$ denote their respective node embedding vectors obtained from the graph embedding model. To calculate network distance, we utilized a dataset of 23,956,566 publications from the PKG dataset published between 1970 and 2019 to construct a MeSH co-occurrence network. The resulting network consists of 29,411 nodes and 45,676,561 edges. We then applied the LINE algorithm to generate a 200-dimensional vector representation for each MeSH term. The dimensionality of the embeddings was set to 200, following the recommendation of Tang et al. [47].

**2) Semantic distance**

To quantify the semantic distance between MeSH terms, this study employs ChatGPT-3.5[4]. a large language model (LLM) recognized for its robust performance in multi-turn dialogue, programming, and text representation. Semantic distance is measured using the model's learned embeddings of MeSH terms, which encode contextual and semantic information derived from real-world textual data [48]. These embeddings ensure that related terms are mapped proximally in the latent space, while unrelated terms remain distant. The semantic distance between MeSH term pairs is computed using Equation (3).

$$d_{semantic}(MeSH_{n-1}, MeSH_n) = 1 - similarity(MeSH_{n-1}, MeSH_n) = 1 - \frac{Sem(MeSH_{n-1})\cdot Sem(MeSH_n)}{||Sem(MeSH_{n-1})||\ ||Sem(MeSH_n)||} \quad (3)$$

Where $Sem(MeSH_{n-1})$ and $Sem(MeSH_n)$ denote embedding vectors of $MeSH_{n-1}$ and $MeSH_n$, respectively, derived from the ChatGPT-3.5 model. To calculate semantic distance, we employed ChatGPT-3.5 to obtain 1,536-dimensional embeddings for 22,200 MeSH terms associated with *PLoS ONE* publications. The embedding dimensionality was set as 1,536 by default and could not be adjusted.

[4] https://chat.openai.com/

**3) Hierarchical distance**

To quantify the hierarchical distance between MeSH terms, we utilized the MeSH term embeddings (named as MeSHHeading2vec) that were trained by Guo et al. [46] based on the MeSH tree structure. Specifically, Guo et al. [46] transformed the MeSH tree into a relational network and employed five graph embedding algorithms to learn vector representations of the terms. Validation results demonstrated that the trained embedding model effectively captures structural information and enhances the representational capacity of the vectors [46]. It should be noted that each MeSH heading may be associated with multiple tree numbers corresponding to different hierarchical positions in the MeSH taxonomy. In the work of Guo et al. [46], each MeSH heading is treated as a unique node in the constructed relationship network. The multiple tree numbers are used only to derive hierarchical relationships by generating parent–child links through iterative truncation of tree number paths. Therefore, the embedding process is performed at the level of MeSH headings, resulting in a single vector representation for each term. The hierarchical distance between MeSH term pairs is computed using Equation (4).

$$d_{hierarchical}(MeSH_{n-1}, MeSH_n) = 1 - similarity(MeSH_{n-1}, MeSH_n) = 1 - \frac{Hie(MeSH_{n-1}) \cdot Hie(MeSH_n)}{||Hie(MeSH_{n-1})||\ ||Hie(MeSH_n)||} \quad (4)$$

Where $Hie(MeSH_{n-1})$ and $Hie(MeSH_n)$ denote embedding vectors of $MeSH_{n-1}$ and $MeSH_n$, respectively, derived from the MeSHHeading2vec. To calculate hierarchical distance, we adopted the 64-dimensional MeSH term embeddings pre-trained by Guo et al. [46]. The embeddings were trained using the MeSH tree structure released in 2019, which includes a total of 29,349 MeSH terms.

It is noteworthy that when the NLM updates the MeSH vocabulary, newly introduced terms are retroactively applied to all previously indexed publications. For example, the term “Heme-Binding Proteins”, introduced in 2020, was retroactively assigned to a 2012 *PLoS ONE* article titled *Heme binding proteins of Bartonella henselae are required when undergoing oxidative stress during cell and flea invasion*. To ensure that all MeSH terms associated with *PLoS ONE* articles had corresponding vector representations, we excluded MeSH terms introduced after 2019, resulting in a final set of 22,183 MeSH terms used in this study.

### 3.2.2 Entropy weight method for comprehensive distances

To quantify comprehensive distances, we applied the Entropy Weight Method [49] (EWM) to determine the weights ($\alpha$, $\beta$, and $1 - \alpha - \beta$, as defined in Equation (1)) for the three distance metrics. This method assigns objective weights based on the dispersion of each metric: a higher dispersion corresponds to a greater weight, and conversely, lower dispersion results in a reduced weight.

We first adopted Min-Max normalization to each distance metric across all MeSH pairs to remove scale differences. For each metric $j$ and MeSH pair $i$, we compute the normalized value $b_{ij}$ and the corresponding proportion $p_{ij}$ using formula (5):

$$p_{ij} = \frac{b_{ij}}{\sum_{i=1}^{m} b_{ij}} \tag{5}$$

Where $m$ is number of MeSH pairs. The entropy of metric $j$ is described as follows:

$$e_j = -k \sum_{i=1}^{m} p_{ij} ln(p_{ij}) \tag{6}$$

Where $k = {}^{1}/_{\ln(m)}$, and we stipulate that $e_j = 0$ when $p_{ij} = 0$. The information entropy redundancy is then defined as $d_j = 1 - e_j$, and the weight of each metric can be expressed as follows:

$$w_j = \frac{d_j}{\sum_{j=1}^{n} d_j} \tag{7}$$

Where $n$ denotes the number of metrics, and weights assigned as $w_1 = \alpha$, $w_2 = \beta$, and $w_3 = 1 - \alpha - \beta$. The comprehensive distance between MeSH pairs is then computed by Equation (1). Using the EWM, the resulting weights were 0.750 for network distance ($\alpha$), 0.073 for semantic distance ($\beta$), and 0.177 for hierarchical distance ($1 - \alpha - \beta$).

### 3.2.3 Novelty measurement of academic papers

The process for calculating a scientific paper's novelty is illustrated in Figure 1, and procedure designed to be implemented as follows:

1) Collect all MeSH terms used in the article and generate all possible term pairs.
2) Using Eq. (2)-(4), compute the network distance, semantic distance, and hierarchical distance for each MeSH pair.
3) Weight the three distance metrics using the result of EWM, then calculate the comprehensive distance for each MeSH pair via Eq. (1).
4) We ranked all MeSH term pairs according to their comprehensive distance in descending order, and defined those in the top 10th percentile as novel combinations, following prior studies on combinatorial novelty[6, 8, 37].
5) For each article, a paper's novelty is quantified by the proportion of novel MeSH pairs to all possible MeSH combinations, and the formula is defined as:

$$novelty_q = \frac{v}{C_p^2} \tag{8}$$

Where $novelty_q$ represents the novelty of article $q$, $v$ denotes the number of novel MeSH pairs in the paper, and $C_p^2$ indicates the total possible MeSH pairs. As illustrated in Figure 1, paper 421456 includes 3 MeSH term pairs, one of which (MeSH_1 and MeSH_2, highlighted in blue) represents a novel combination. Consequently, the paper's novelty score is computed

as 1/3.

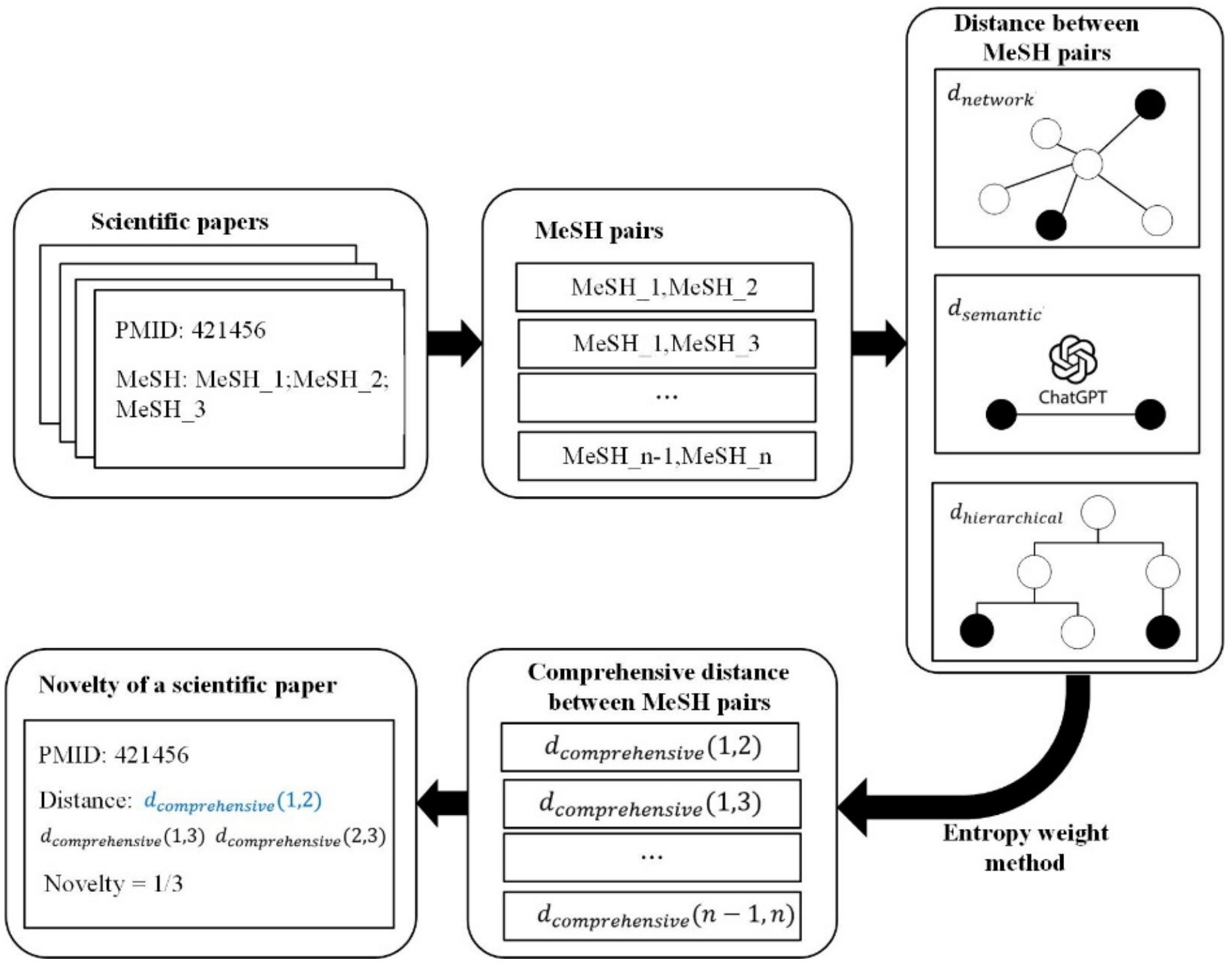

Figure 1 The process for measuring the novelty of scientific papers

## 3.3 Evaluation of novelty measurement

Inspired by Bornmann et al. [43], we compared the novelty score of papers with and without novelty tags, as described in Table 2. The underlying rationale is that papers explicitly labeled as novel should, on average, exhibit higher novelty scores than those without such tags. For each paper with novelty tags, we identified a matched counterpart (also referred to as control group) from *PLoS ONE* papers without novelty tags to enhance comparability between the two groups. The matched paper was required to meet the following criteria: it had to be published in the same year, volume, issue, and discipline, and have the same number of authors. When multiple candidates satisfied these conditions, one was randomly selected. To ensure the robustness of our findings, we repeated the matching procedure five times.

Table 3 presents the matching results between *PLoS ONE* papers from the H1 Connect platform and their control group. “Positive label” indicates that the recommended *PLoS ONE* paper was tagged with one or more of the following: “Interesting Hypothesis”, “New Finding”, “New Drug Target”, or “Technical Advance”. Notably, some articles received conflicting labels from different experts. For instance, a single paper could be labeled both “Confirmation” and “Interesting Hypothesis”. To mitigate inconsistencies in expert evaluations, such cases were excluded from the matching process.

Table 3 The matching results between *PLoS One* papers from the H1 Connect platform and their control group

| **Category** | **Positive Label** | **Interesting Hypothesis** | **New Finding** | **New Drug Target** | **Technical Advance** |
|---|---|---|---|---|---|
| Papers recommended by H1 connect | 874 | 188 | 764 | 84 | 206 |
| Control group papers | 874 | 188 | 764 | 84 | 206 |

**Note**: The table displays outcomes from one matching iteration.

During the validation, we employed least squares regression to estimate the effect of novelty tags on novelty scores. Specifically, we treated the novelty score as the dependent variable and used a binary indicator as the independent variable (coded as 1 for papers with novelty tags and 0 for those without), allowing us to examine the relationship between the novelty indicators and expert evaluations.

# 4. Results

## 4.1 Comparative analysis of the three distance metrics

To compare the distribution of MeSH terms under different distance metrics, we applied Uniform Manifold Approximation and Projection (UMAP) to visualize the MeSH term vectors in two-dimensional space and clustered formats [50]. The main parameters were set as *n_neighbors* = 100 and *min_dist* = 0.1. The results are shown in Figure 2, where each node denotes a MeSH term, and colors represent its classification into one of 16 MeSH categories[5]. The visualization reveals distinct distribution patterns corresponding to the three types of distance metrics. The embeddings based on the MeSH tree structure exhibit relatively well-defined clusters (Figure 2C), reflecting their alignment with expert-curated hierarchical classifications. In contrast, the embeddings generated by ChatGPT (Figure 2B) exhibit a noticeably greater degree of cluster overlap. Furthermore, the embeddings derived from the co-occurrence network (Figure 2A) display more intermixed clusters, which may reflect less clear structural separation. These observations suggest that each distance metric captures different latent dimensions of the relationships between MeSH terms.

[5] https://meshb.nlm.nih.gov/treeView[50]

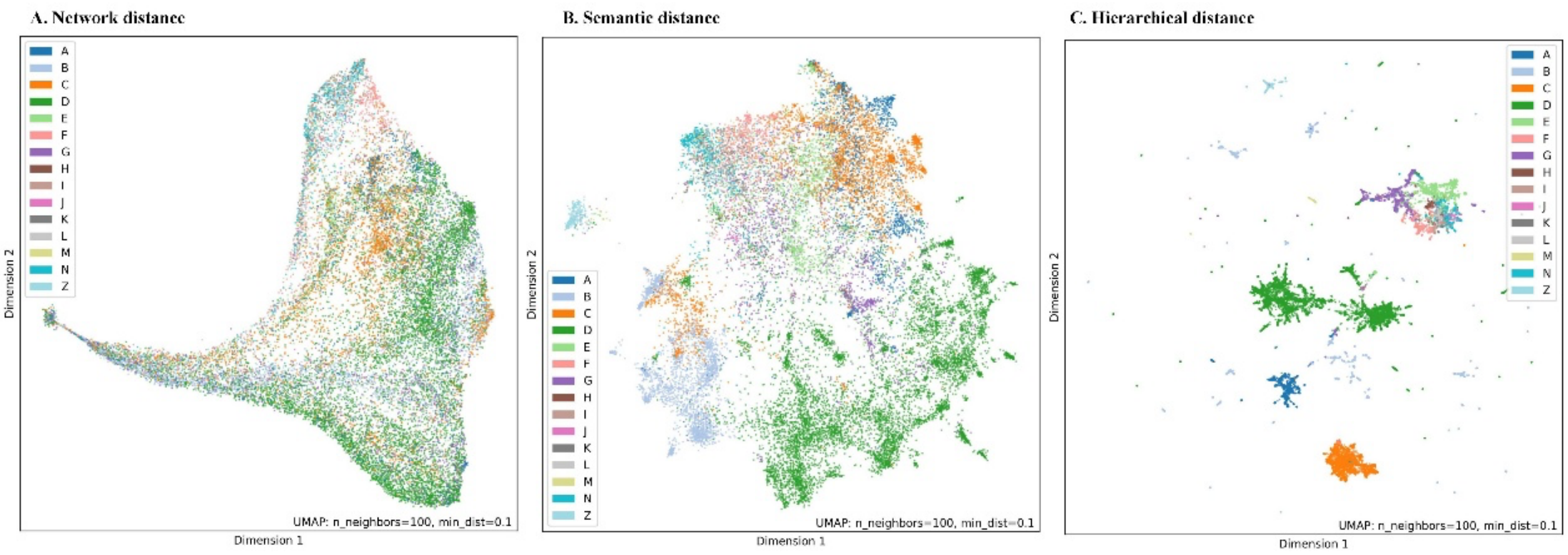

Figure 2 The UMAP visualization of MeSH terms.

To quantitatively assess the differences among the three distance metrics, we applied a series of statistical analyses. First, we computed the distances between MeSH term pairs using three distinct metrics: network-based, semantic, and hierarchical. To compare the distributions of the resulting distance values, we employed the Friedman test [51], which revealed extremely significant differences among the three groups ( $\chi^2(2) = 2933225.76, p < 0.001$ ). To examine differences between each pair of distance metrics, we further conducted Wilcoxon signed-rank tests and applied Bonferroni correction to control for multiple testing [52]. As shown in Table 4, the network distance differed significantly from both the semantic distance ($Z = -1484.38, p < 0.001, r = -0.82$) and the hierarchical distance ($Z = -1203.24, p < 0.001, r = -0.66$). Additionally, the difference between semantic and hierarchical distances was also statistically significant ($Z = 881.75, p < 0.001, r = -0.49$). According to Cohen's [53] guidelines for interpreting effect sizes in non-parametric tests, the absolute values of the effect sizes indicate large effects for the first two comparisons ($|r| > 0.5$), and a medium effect for the third comparison ($0.2 < |r| < 0.5$). These findings suggest substantial and meaningful differences across all three types of distance metrics.

Table 4 The result of pairwise comparisons among three distance metrics

| Group | Z value | P value(adj) | Effect size(r) |
|---|---|---|---|
| Network distance vs. Semantic distance | -1484.38 | 0.001 | -0.82 |
| Network distance vs. Hierarchical distance | -1203.24 | 0.001 | -0.66 |
| Semantic distance vs. Hierarchical distance | 881.75 | 0.001 | 0.49 |

Note: P value(adj) represents the P value with Bonferroni correction.

## 4.2 Validation results of novelty measurement

### 4.2.1 Descriptive analysis of novelty score

Table 5 reports the descriptive statistics of novelty scores for papers recommended by H1 Connect and their corresponding control group papers. On average, papers with positive labels and recommended by the H1 Connect platform have a mean novelty score of 0.138, which is slightly higher than the control group's mean of 0.135. The median novelty score is also higher for the recommended papers, reaching 0.130, while the control group papers have a median of 0.128. Across specific novelty tags, such as "Interesting Hypothesis", "New Finding", "New Drug Target", and "Technical Advance", the recommended papers generally show higher mean novelty scores than their counterparts. A similar pattern is observed in the median scores, except for papers labeled "Technical Advance", which do not display a clear advantage. Overall, these results are consistent with our expectations and provide indirect support for the validity of the novelty measurement.

Table 5 The descriptive statistics of novelty scores

| **Category** | **Descriptive Statistics** | **Positive Label** | **Interesting Hypothesis** | **New Finding** | **New Drug Target** | **Technical Advance** |
|---|---|---|---|---|---|---|
| Papers recommended by H1 connect | Mean | 0.138 | 0.157 | 0.141 | 0.156 | 0.135 |
| | Median | 0.130 | 0.150 | 0.133 | 0.142 | 0.123 |
| | Std | 0.096 | 0.105 | 0.096 | 0.083 | 0.104 |
| | Min | 0 | 0 | 0 | 0 | 0 |
| | Max | 0.5 | 0.667 | 0.5 | 0.352 | 0.436 |
| | N | 4370 | 940 | 3820 | 420 | 1030 |
| Control group papers | Mean | 0.135 | 0.135 | 0.135 | 0.141 | 0.131 |
| | Median | 0.128 | 0.128 | 0.127 | 0.133 | 0.124 |
| | SD | 0.090 | 0.092 | 0.092 | 0.084 | 0.095 |
| | Min | 0 | 0 | 0 | 0 | 0 |
| | Max | 0.6 | 0.667 | 0.7 | 0.5 | 1 |
| | N | 4370 | 940 | 3820 | 420 | 1030 |

### 4.2.2 Validation results via linear regression modeling

To further assess the validity of our novelty measure, we adopted a linear regression model to examine whether the proposed indicator aligns with peer judgements. Specifically, we tested whether the average novelty scores differ significantly between papers labeled with novelty-related tags and those without such tags. Additionally, we compared the performance of our text-based novelty measure with the reference-based novelty measure developed by Uzzi et al. [6]. The validation results are presented in Table 6. Remarkably, our novelty metric is a positive indicator, whereas the novelty score U is a negative indicator. Accordingly, we expected our novelty scores to be positively associated with novelty-related tags, while the novelty score U would exhibit a negative association.

Table 6 The validation result using regression models

| Variable | Our novelty score | | | | |
|---|---|---|---|---|---|
| | (1) | (2) | (3) | (4) | (5) |
| Positive Label | 0.003<br>(0.002) | | | | |
| Interesting Hypothesis | | 0.022***<br>(0.005) | | | |
| New Finding | | | 0.006***<br>(0.002) | | |
| New Drug Target | | | | 0.015***<br>(0.006) | |
| Technical Advance | | | | | 0.004<br>(0.004) |
| Constant | 0.135***<br>(0.001) | 0.135***<br>(0.000) | 0.135***<br>(0.000) | 0.141***<br>(0.000) | 0.131***<br>(0.000) |
| N | 8740 | 1880 | 7640 | 840 | 2060 |
| | **Novelty score U** | | | | |
| | (6) | (7) | (8) | (9) | (10) |
| Positive Label | 0.345<br>(0.440) | | | | |
| Interesting Hypothesis | | -1.601*<br>(0.850) | | | |
| New Finding | | | 0.814*<br>(0.442) | | |
| New Drug Target | | | | -2.444***<br>(0.847) | |
| Technical Advance | | | | | -1.097<br>(0.966) |
| Constant | -2.897***<br>(0.302) | -3.188***<br>(0.529) | -2.948***<br>(0.069) | -3.992***<br>(0.678) | -2.491***<br>(0.529) |
| N | 8740 | 1880 | 7640 | 840 | 2060 |

**Note:** Robust standard errors in parentheses. *, **, and *** denote significance at the 10%, 5%, and 1% level, respectively.

As shown in Table 6, our novelty scores exhibit positive coefficients for all novelty tags. However, the coefficients for the tags “Positive Label” and “Technical Advance” are not statistically significant. Interestingly, as discussed in Section 4.3, when novelty is measured using the 5th-threshold specification, which applies a more stringent criterion for identifying novel combinations, all novelty tags show significant positive associations with the novelty scores. One possible explanation is that different novelty tags may correspond to different levels of expectation violation [54]. Compared with tags such as “Interesting Hypothesis” or “New Drug Target,” the label “Technical Advance” may represent a higher threshold of novelty. In other words, a study may need to incorporate sufficiently rare and atypical combinations of knowledge elements before it is recognized as a genuine technical advance.

In contrast, the results for novelty score U are less favorable. Among the four specific novelty tags, only “Interesting Hypothesis,” “New Drug Target,” and “Technical Advance” exhibit coefficients in the expected direction, and only the first two are statistically significant. Furthermore, although novelty score U is significantly associated with the “New Finding” tag, the direction of the relationship is contrary to expectations.

Overall, the results for our novelty score are mostly consistent with our expectations regarding the relationship between novelty and expert-assigned tags. In other words, the proposed novelty measure aligns well with expert judgment, supporting its validity. Compared to Uzzi’s et al. [6] measure, our measure provides a more valid and reliable reflection of scientific novelty.

### 4.2.3 Effectiveness of combining different distance metrics in novelty measurement

To further verify whether integrating multiple distance metrics improves the accuracy of scientific novelty quantification, we conducted comparative experiments between novelty measures derived from combined metrics and those based on individual metrics. Following the methodology described in Section 3.2, we computed three distinct novelty measures based on network, semantic, and hierarchical distance separately. Similarly, using the evaluation method outlined in Section 3.3, we estimated the correlation between novelty scores and novelty-related tags. The validation results are shown in Figure 3, and the detailed corresponding regression results are provided in Appendix Table A.1.

In Figure 3, we can observe that only a small number of coefficients are positive and significant, aligning with our expectations. Regarding novelty scores derived from network distance, only two tags, “Interesting Hypothesis” and “New Drug Target” meet these expectations. For novelty scores based on semantic distance, the coefficients for “Interesting Hypothesis” and “New Finding” point in the expected direction, but neither reaches statistical significance. For novelty scores based on hierarchical distance, the tags “Positive Label, “Interesting Hypothesis”, “New Finding” and “New Drug Target” all display positive associations. However, only the coefficient for “New Drug Target” is statistically significant.

Taken together, novelty scores calculated from individual distance metrics align with only a few of our previously formulated expectations. In contrast, novelty scores derived from combined distance metrics show better agreement with novelty-related tags recommended by peers, which implies that integrating different distance measures enhances the ability to characterize the scientific novelty of academic papers.

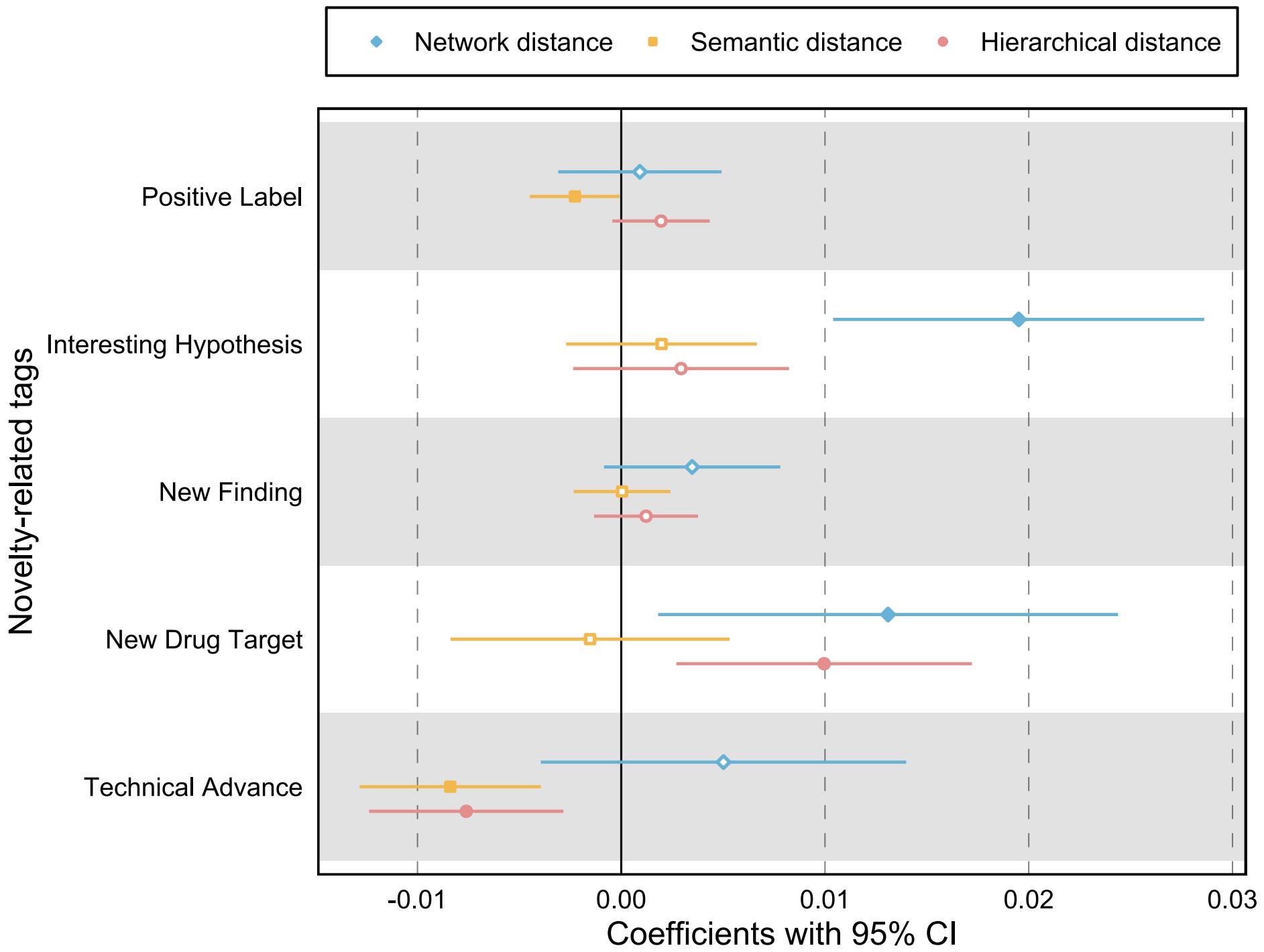

Figure 3 Validation of novelty measurements developed using separate distance metrics. Whiskers represent 95% confidence intervals while solid and hollow geometric shapes denote significant and non-significant coefficients, respectively. Colors denote novelty scores based on different distance metrics: blue diamonds correspond to network distance, orange squares to semantic distance, and red circles to hierarchical distance.

## 4.3 Robustness tests

In this section, we examine the robustness of our novelty measures. First, although we adopt the 10th percentile of the comprehensive distance as the threshold to distinguish novel from common knowledge combinations, which is consistent with prior studies [6, 8, 17], the choice of threshold may influence the resulting novelty scores of academic papers. To address this concern, we conduct sensitivity analyses using alternative thresholds at the 5th and 15th percentiles. The corresponding results are reported in Tables A.2 and A.3. As shown in Table A.2, all novelty tags show a positive and statistically significant correlation with our novelty score. Table A.3 presents similar results, and all novelty tags remain positively correlated with the novelty score, although only the coefficients for “Interesting Hypothesis” and “New Drug Target” are statistically significant. Overall, the signs of the coefficients are consistent with those reported in Table 6, while differences in statistical significance are minor. These differences are reasonable, as the percentile threshold can be interpreted as a hyperparameter that determines the operational definition of novelty. Lower thresholds impose a stricter criterion, capturing only highly atypical MeSH term combinations, whereas higher thresholds relax this constraint and include a broader set of more common MeSH term combinations. This, in turn, leads to varying degrees of alignment with novelty tags assigned by human experts. The results further suggest that lower thresholds are more closely aligned with expert judgments. Second, to examine the sensitivity of our novelty measure to the weighting scheme, we constructed an alternative novelty measure by assigning equal weights to the network, semantic, and hierarchical distance dimensions instead of using EWM-derived weights. As reported in Table A.4, the conclusions remain unchanged relative to the main analysis (Table 6), suggesting that our novelty measure is robust to alternative weighting schemes.

## 4.4 Case study

To enhance the interpretability of the proposed novelty measure, we selected two articles from radiology and nuclear medicine, which is a subdiscipline of clinical medicine, to demonstrate how our method distinguishes the novelty between these two articles, as shown in Figure 4. Figure 4A presents a low-novelty article tilted *Robust framework for PET image reconstruction incorporating system and measurement uncertainties*, which is indexed with six MeSH terms. Several of these terms are semantically related and belong to the same major branches (i.e., Category L: Information Science) of the MeSH tree, including "Algorithms", "Computer Simulation", "Image Processing, Computer-Assisted", and "Models, Theoretical". These relationships result in relatively short distances between the MeSH term pairs. Consequently, none of the fifteen MeSH term pairs exceed the threshold value for comprehensive distance and are therefore identified as novel combinations, resulting in a novelty score of zero.

In contrast, the high-novelty article shown in Figure 4B, titled *Estimation of noise-free variance to measure heterogeneity*, is indexed with four MeSH terms. These MeSH terms co-occur with relatively low frequency and are generally located in different branches of the MeSH tree, resulting in comparatively large network and hierarchical distances. For example, the MeSH term "Humans" belongs to the Category B (Organisms), whereas "Image Processing, Computer-Assisted" falls under Category L (Information Science), displaying far hierarchical distance. Moreover, according to our statistics, "Humans" and "Image Processing, Computer-Assisted" co-occur only twice in our dataset, implying a large network distance. As a result, four of the six MeSH term pairs are identified as novel combinations, leading to a much higher novelty score. These contrasting examples illustrate how the proposed method captures differences in conceptual novelty by examining multidimensional distances between the MeSH terms assigned to individual articles.

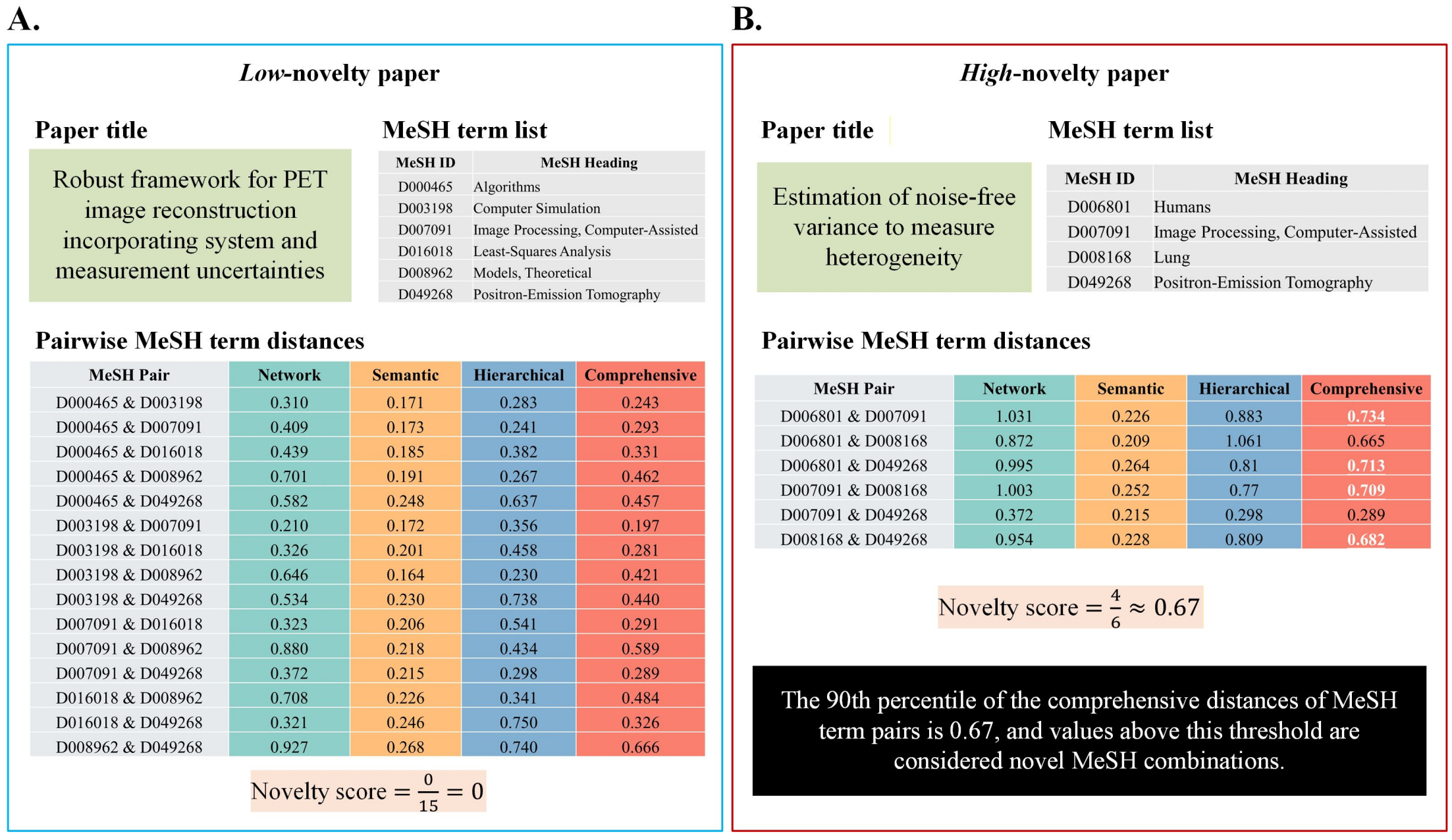

Figure 4 Novelty calculation process for two articles

# 5. Discussion

In this section, we will discuss the implications and limitations of our study.

## 5.1 Theoretical implications

Novelty plays a crucial role in innovative scientific publications, and establishing a reliable and accurate method for quantifying novelty holds great significance for the scientific community [55]. From a combinatorial perspective, numerous novelty measures have been developed, ranging from metadata-based to content-based approaches [12, 54]. Content reflects how ideas are described and expressed within disciplines. Therefore, this study treats MeSH terms as knowledge units and proposed a novel content-based approach to quantify the novelty of scientific papers from a combinatorial angle, considering three latent distances: network, semantic, and hierarchical distances between knowledge units.

The main theoretical contribution of this study is the introduction of a comprehensive distance metric between knowledge units for measuring scientific novelty. This study hypothesizes that distinct distance measures can capture different relationships between knowledge units in latent space. The statistical results show that the three types of distance metrics differ significantly from one another, and integrating these metrics offers clear advantages in measuring scientific novelty compared to using the three separate novelty measures. This novelty measure, based on a comprehensive distance metric, expands the scope of combinatorial novelty research and provides a more reliable quantification of a paper's novelty, avoiding conceiving the process of scientific innovation in a simplified way.

Although our novelty measure draws on MeSH terms from the biomedical domain, it exhibits strong generalizability comparable to other metadata-based approaches. McNamee [18] argued that, to understand the complex phenomenon in the world, entities are often categorized and organized within detailed taxonomic systems. Classification frameworks in science and technology are no exception and typically feature intricate hierarchical architectures. Notable examples includes Dewey Decimal Classification (DDC) [56], the International Patent Classification (IPC) scheme [57], fields of study (FOS) developed by Microsoft Academic Graph [58], Computer Science Ontology (CSO) curated by SKM team [59], as well as Gene Ontology (GO) maintained by gene ontology consortium [60]. Therefore, following our methodology, novelty assessment can be extended to other domains that utilize established hierarchical classification systems, enabling cross-disciplinary comparisons and broadening the applicability of our measure beyond biomedicine.

## 5.2 Practical implications

This study has several practical implications as well. First, although integrating multiple distance metrics to assess the scientific novelty of academic papers has shown promising results, the resulting novelty scores do not fully align with the peer judgements. This discrepancy indicates that automated evaluations cannot replace peer review, which remains the cornerstone of scientific assessment. As stated by Thelwall et al. [61], automatic methods for evaluating novelty in academic research still require further development.

Second, novelty assessment is a crucial component of research quality and should be distinguished from the evaluation of scientific impact. Current evaluation practices for journals, researchers, and institutions rely heavily on metadata-based indicators such as publication counts, citation metrics, and related measures. In this context, our novelty indicator can serve as a valuable supplement to the existing evaluation framework. As many scholars have argued, a robust scientific assessment system should evaluate entities from multiple perspectives and incorporate diverse indicators to ensure a comprehensive, objective, and balanced appraisal [62, 63].

Third, academic search engines can also benefit from our research. Most existing platforms rank search results based on relevance or publication date, which often falls short of helping scholars quickly identify cutting-edge research. By incorporating our novelty indicator and offering an option to sort results by novelty, search engines can significantly enhance the user experience for academic researchers.

### 5.3 Limitations

This study also has several limitations. Firstly, this study validated the proposed novelty measure using academic papers from the biomedical domain. Whether the measure maintains its effectiveness across other fields remains to be further explored. Secondly, as a multidimensional construct, novelty can be interpreted from various perspectives. While our measure focuses on the combinatorial aspect, other dimensions, such as uniqueness, bridging, and surprise, may also offer valuable bases for measuring novelty. Thirdly, although MeSH is an expert-curated and standardized thesaurus that offers advantages for measuring scientific novelty, its relatively low update frequency may lead to an underestimation of the novelty of some papers. Fourth, the representation of each MeSH term as a single node, despite its multiple tree-number assignments, may partially obscure branch-specific hierarchical information and may influence the fine-grained assessment of scientific novelty within the distance-based evaluation framework. Fifth, we employ a global threshold to capture absolute novelty across the entire dataset, which may overlook temporal heterogeneity in knowledge combination structures, as novelty distributions may vary across publication years due to the evolving knowledge base. Sixth, the network distance measure is derived from a MeSH co-occurrence network constructed using publications from 1970 to 2019, whereas the focal articles were published between 2007 and 2015. As a result, post-publication relationships among knowledge elements may be incorporated into the measure, potentially introducing temporal leakage and attenuating estimates of novelty. Future research could address this issue by constructing year-specific knowledge networks to better capture the historical context in which scientific contributions were made.

## 6. Conclusion and future work

In this study, we propose a novel approach to measuring the scientific novelty of academic papers from a recombination perspective. Unlike existing keyword- or entity-based measures that typically reflect a single dimension of knowledge recombination, our method incorporates multiple complementary dimensions (i.e., network distance, semantic distance, and hierarchical distance) between knowledge units. We use MeSH terms as proxies for knowledge units and validate the approach using articles from PLOS ONE alongside novelty tags from the H1 Connect platform. The results indicate that these three dimensions capture distinct aspects of the relationships among MeSH terms, and that integrating them offers clear advantages in identifying novel papers compared to using any single indicator alone.

This study also leaves several avenues for future research. First, the proposed novelty measure could be applied to other disciplines to assess its effectiveness and generalizability. A key prerequisite for such validation is the construction of a larger, more reliable, and unbiased ground-truth dataset. Second, further theoretical exploration into the nature of scientific novelty, its origins, dimensions, and underlying mechanisms, would greatly contribute to the refinement, simulation, and generation of novelty in quantitative studies. Third, future work could adopt hypergraph models to capture higher-order combinations of knowledge units, offering a more nuanced understanding of complex recombination patterns and their role in the emergence of novelty. Fourth, the MeSH thesaurus represents a type of human knowledge, and integrating human knowledge with global knowledge such as that encoded in large language models provides an interesting pathway for quantifying scientific novelty [64].

## Acknowledgements

This work was supported by the National Social Science Fund of China (No. 24CTQ027). The current study is an extended version of our article [15], which was presented at the 2024 ACM/IEEE Joint Conference on Digital Libraries held in Hongkong, China, from December 16 to 20, 2024. The authors would like to thank the two anonymous reviewers for their insightful comments and valuable suggestions.

# Appendix

Table A.1 Validation results of novelty measures developed with distinct distance metrics

| **Variable** | **Network-distance-based novelty score** | | | | |
|---|---|---|---|---|---|
| | **(1)** | **(2)** | **(3)** | **(4)** | **(5)** |
| Positive Label | 0.0009<br>(0.0020) | | | | |
| Interesting Hypothesis | | 0.0195***<br>(0.0046) | | | |
| New Finding | | | 0.0035<br>(0.0022) | | |
| New Drug Target | | | | 0.0131**<br>(0.0058) | |
| Technical Advance | | | | | 0.0050<br>(0.0046) |
| Constant | 0.1424***<br>(0.0014) | 0.1432***<br>(0.0032) | 0.1419***<br>(0.0015) | 0.1468***<br>(0.0042) | 0.1390***<br>(0.0031) |
| N | 8740 | 1880 | 7640 | 840 | 2060 |
| **Variable** | **Semantic -distance-based novelty score** | | | | |
| | **(6)** | **(7)** | **(8)** | **(9)** | **(10)** |
| Positive Label | -0.0023**<br>(0.0011) | | | | |
| Interesting Hypothesis | | 0.0020<br>(0.0024) | | | |
| New Finding | | | 0.0000<br>(0.0012) | | |
| New Drug Target | | | | -0.0015<br>(0.0035) | |
| Technical Advance | | | | | -0.0084***<br>(0.0023) |
| Constant | 0.0501***<br>(0.0008) | 0.0476***<br>(0.0017) | 0.0493***<br>(0.0009) | 0.0563***<br>(0.0026) | 0.0507***<br>(0.0017) |
| N | 8740 | 1880 | 7640 | 840 | 2060 |
| **Variable** | **Hierarchical -distance-based novelty score** | | | | |
| | **(11)** | **(12)** | **(13)** | **(14)** | **(15)** |
| Positive Label | 0.0020<br>(0.0012) | | | | |
| Interesting Hypothesis | | 0.0029<br>(0.0027) | | | |
| New | | | 0.0012 | | |

| Variable | (1) | (2) | (3) | (4) | (5) |
|---|---|---|---|---|---|
| Finding | | | (0.0013) | | |
| New Drug Target | | | | 0.0100***<br>(0.0037) | |
| Technical Advance | | | | | -0.0076***<br>(0.0024) |
| Constant | 0.0946***<br>(0.0009) | 0.0945***<br>(0.0020) | 0.0956***<br>(0.0009) | 0.0994***<br>(0.0027) | 0.0941***<br>(0.0018) |
| N | 8740 | 1880 | 7640 | 840 | 2060 |

**Note:** Robust standard errors in parentheses. *, **, and *** denote significance at the 10%, 5%, and 1% level, respectively.

Table A.2 Validation results of novelty measures under the 5th percentile threshold

| Variable | (1) | (2) | (3) | (4) | (5) |
|---|---|---|---|---|---|
| Positive Label | 0.003**<br>(0.001) | | | | |
| Interesting Hypothesis | | 0.018***<br>(0.003) | | | |
| New Finding | | | 0.005**<br>(0.002) | | |
| New Drug Target | | | | 0.012***<br>(0.004) | |
| Technical Advance | | | | | 0.006**<br>(0.004) |
| Constant | 0.074***<br>(0.001) | 0.075***<br>(0.002) | 0.074***<br>(0.001) | 0.077***<br>(0.003) | 0.074***<br>(0.002) |
| N | 8740 | 1880 | 7640 | 840 | 2060 |

**Note:** Robust standard errors in parentheses. *, **, and *** denote significance at the 10%, 5%, and 1% level, respectively.

Table A.3 Validation results of novelty measures under the 15th percentile threshold

| Variable | (1) | (2) | (3) | (4) | (5) |
|---|---|---|---|---|---|
| Positive Label | 0.001<br>(0.002) | | | | |
| Interesting Hypothesis | | 0.024***<br>(0.005) | | | |
| New Finding | | | 0.004<br>(0.003) | | |
| New Drug Target | | | | 0.013**<br>(0.007) | |
| Technical Advance | | | | | 0.003<br>(0.005) |
| Constant | 0.196***<br>(0.002) | 0.197***<br>(0.004) | 0.197***<br>(0.002) | 0.208***<br>(0.005) | 0.189***<br>(0.004) |
| N | 8740 | 1880 | 7640 | 840 | 2060 |

**Note:** Robust standard errors in parentheses. *, **, and *** denote significance at the 10%, 5%, and 1% level, respectively.

Table A.4 Validation results for the equal-weight novelty measure (10th Percentile Threshold)

| **Variable** | (1) | (2) | (3) | (4) | (5) |
|---|---|---|---|---|---|
| Positive Label | 0.003<br>(0.002) | | | | |
| Interesting Hypothesis | | 0.016***<br>(0.004) | | | |
| New Finding | | | 0.007***<br>(0.002) | | |
| New Drug Target | | | | 0.018***<br>(0.005) | |
| Technical Advance | | | | | 0.002<br>(0.004) |
| Constant | 0.109***<br>(0.001) | 0.109***<br>(0.003) | 0.108***<br>(0.001) | 0.116***<br>(0.004) | 0.108***<br>(0.002) |
| N | 8740 | 1880 | 7640 | 840 | 2060 |

**Note:** Robust standard errors in parentheses. *, **, and *** denote significance at the 10%, 5%, and 1% level, respectively.